\documentclass[12pt]{iopart}

\usepackage{graphicx}

\begin{document}

\title{The Strange Prospects for Astrophysics}

\author{I Sagert\dag,
M Hempel\dag,
G Pagliara\ddag,
J Schaffner-Bielich\ddag\footnote[5]{invited talk given at the
    International Conference on Strangeness in Quark Matter (SQM2008),
    Beijing, China, October 6-10, 2008.}}

\address{
\dag\ Institut f\"ur Theoretische Physik, Goethe Universit\"at, 
Max-von-Laue-Str.~1, 60438~Frankfurt am Main, Germany\\
\ddag\ Institut f\"ur Theoretische Physik, Ruprecht-Karls-Universit\"at, Philosophenweg 16, 69120 Heidelberg, Germany}

\author{T Fischer\S,
A Mezzacappa$\|$,
F-K Thielemann\S\ 
and
M~Liebend\"orfer\S}

\address{
\S\ Department of Physics, University of Basel, Klingelbergstr.~82, 4056 Basel,           
Switzerland \\
$\|$\ Physics Division, Oak Ridge National Laboratory, Oak Ridge, TN 37831, USA}

\ead{schaffner@thphys.uni-heidelberg.de}

\begin{abstract}
  The implications of the formation of strange quark matter in neutron
  stars and in core-collapse supernovae is discussed with special
  emphasis on the possibility of having a strong first order QCD phase
  transition at high baryon densities. If strange quark matter is
  formed in core-collapse supernovae shortly after the bounce, it
  causes the launch of a second outgoing shock which is energetic
  enough to lead to a explosion. A signal for the formation of strange
  quark matter can be read off from the neutrino spectrum, as a second
  peak in antineutrinos is released when the second shock runs over
  the neutrinosphere.
\end{abstract}



\section{Introduction}


The exploration of the QCD phase diagram is not only a task for
heavy-ion physics as there are strong relations to astrophysics of
extremely dense matter and even to cosmology. The conditions in the
early universe are similar to the ones probed at the heavy-ion
collisions at RHIC and LHC, high temperatures and low net baryon
densities. Supernova matter and neutron star matter is located in the
QCD phase diagram at moderate temperatures and high net baryon
densities. In this region of the QCD phase diagram one expects to have
a strong first order phase transition which is related to the
restoration of chiral symmetry%
\footnote{In the following, we will refer to the new phase at high
  densities generically to 'quark matter' although matter is not
  necessarily deconfined as the phase transition is related to chiral
  symmetry restoration.}.  QCD matter at extreme baryon densities will
be investigated by heavy-ion experiments with FAIR at GSI Darmstadt.

\begin{figure}
\begin{center}
\includegraphics[width=0.6\textwidth]{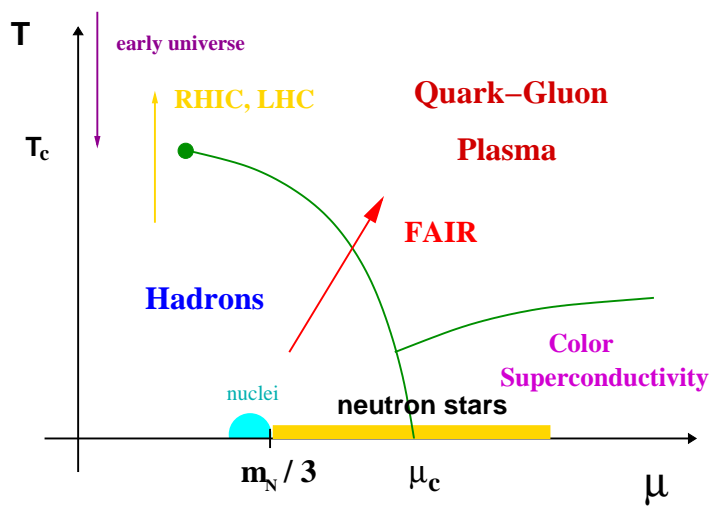}
\caption{The QCD phase diagram, the lines denote first order phase
  transition which are due to chiral symmetry breaking and/or the
  formation of color superconducting quark matter (taken from
  \cite{SchaffnerBielich:2007mr}).}
\label{fig:qcdphase}
\end{center}
\end{figure}

The QCD equation of state (EoS) is an essential input in astrophysical
systems.  In core-collapse supernovae simulations temperatures of
about $T=10-20$ MeV with densities slightly above normal nuclear
matter density are reached at core bounce, the newly born
proto-neutron star is heated up to $T=50$ MeV with densities a few
times normal nuclear matter densities, the final cold neutron stars
then has central densities of up to ten times normal nuclear matter
densities for a soft equation of state. Finally, neutron star mergers
simulations can achieve temperatures of typically $T=30$ MeV, even
higher temperatures have been noticed for some equations of
states. Note, that the dynamical timescales involved are usually not much
less than around a few 100 microseconds so that they are always much
larger than the timescale to establish equilibrium with respect to weak interactions involving
strange particles of about $10^{-8}$ seconds or less. Hence, the
matter for astrophysical applications is necessarily in
weak equilibrium with respect to strangeness and always includes
strange matter! Except for cold neutron stars, there is a subtlety
here, as protons and neutrons are not in weak equilibrium. Hence, for
dynamical astrophysical scenarios, the matter is characterized by a
given temperature $T$, baryon density $n$ and proton fraction $Y_p$.

For hunting down strange quark matter in the heavens several signals
have been suggested in the literature as exotic mass-radius relation
of compact stars, rapidly rotating pulsars due to r-mode {\em
  stability} window, enhanced cooling of neutron stars, and gamma-ray
bursts by transition to strange quark matter. Let us first concentrate
on strange quark matter in neutron star before we discuss the
implications of the formation of strange quark matter for
core-collapse supernovae.


\section{Strange Quark Matter in Neutron Stars}


Neutron stars are produced in core-collapse supernova explosions and
are extremely compact, massive objects with radii of $\approx$ 10 km
and masses of $1-2M_\odot$, involving extreme densities, several times
nuclear density: $n\gg n_0 = 0.16$~fm$^{-3}$.

More than 1700 pulsars, rotation-powered neutron stars, are presently
known.  The best determined mass is the one of the Hulse-Taylor
pulsar, $M=(1.4414\pm 0.0002)M_\odot$ \cite{Weisberg:2004hi}, the
smallest known mass is $M=(1.18\pm 0.02)M_\odot$ for the pulsar
J1756-2251 \cite{Faulkner:2005}. Note, that the mass of the pulsar
J0751+1807 has been corrected from $M=2.1\pm 0.2 M_\odot$ to
$M=1.14-1.40M_\odot$\cite{Nice:2008}.  We note that the extremely
large neutron star masses extracted from pulsars found in globular
clusters \cite{Freire:2007jd} can only give an upper bound. Only the
periastron advance of the pulsar PSR J1748-2021B has been determined
so far, the inclination angle $i$ of the orbital plane is still
unknown. A statistical analysis for that angle is not really
appropriate for one pulsar.  For an inclination angle of $i=4-5$
degrees, two neutron stars with a mass of $M\sim 1.4M_\odot$ are
possible.  A measurement of a second effect from general relativity is
needed to draw a firm final conclusions. 

The spectral analysis of the closest isolated neutron star known, RX
J1856.5-3754, hints at more complex surface properties than initially
expected.  Fits with a two-component blackbody as well as with a
condensed surface and a small layer of hydrogen result in rather large
radiation radii $R_\infty = R/\sqrt{1-2GM/R} = 17 (d/140 pc)$~km. With
an inferred gravitational redshift of $z_g\approx 0.22$, the neutron
star would have a true radius of $R\approx 14$ km and a mass of
$M\approx 1.55M_\odot$ \cite{Ho:2006uk}. A large uncertainty
resides in the still not well known distance $d$, but clearly more
data and analysis is needed to understand the atmosphere and the
radiation of neutron stars.

In binary systems of a neutron star with an ordinary star accreting
material falling onto the neutron star ignites nuclear burning which
is observable as an x-ray burster.
The analysis of \"Ozel \cite{Ozel:2006km} for the x-ray burster EXO
0748--676 arrived at mass-radius constraints of $M\geq 2.10\pm 0.28
M_\odot$ and $R\geq 13.8 \pm 1.8$ km. The values derived have to be
taken with great care, as a multiwavelength analysis of Pearson et
al.\ \cite{Pearson:2006zy} concludes that the data is more consistent
with a mass of $M=1.35M_\odot$ than with $M=2.1M_\odot$.  Even if such
large masses and radii are taken for granted, quark matter can still
be present in the interior of neutron stars as demonstrated by Alford
et al.\ \cite{Alford:2006vz}. The limits would rule out soft equations
of states, not quark stars or hybrid stars, compact stars with a
hadronic mantle and a quark matter core.

Future telescopes and detectors will probe compact stars in more
details as the International X-ray Observatory IXO, the James Webb
Space Telescope JWST, the Square Kilometre Array SKA, LISA and UNO, an
underground neutrino observatory.  With the future x-ray satellites
one can measure the profile of the burst oscillations which is
modified from the space-time warpage around the compact star.  By this
method a model independent measurement of the mass and radius of the
compact star can be extracted. It was claimed that one could
determine the mass-radius ratio to within 5\% with Constellation-X with this method \cite{Bhattacharyya:2005ge}.

\begin{figure}
\begin{center}
\includegraphics[width=0.6\textwidth]{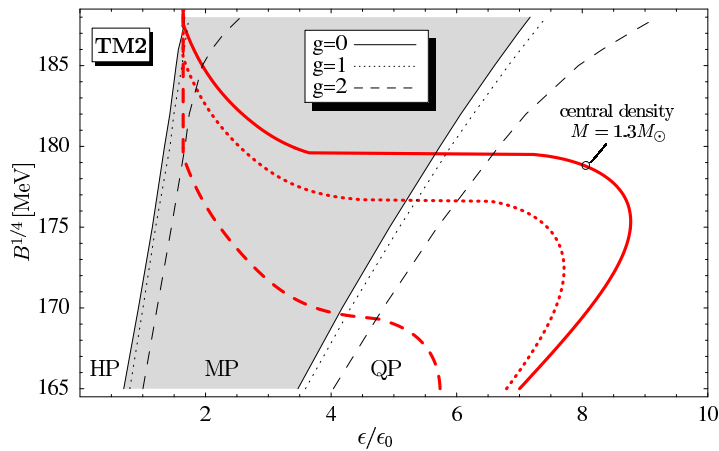}  
\caption{The phase structure of hybrid stars within the MIT bag model
  and using a HDL approximation for different values of the MIT bag
  constant. HP: Hadronic phase, MP: mixed phase, QP: Quark phase.
  Reprinted from \cite{Schertler:2000xq}, Copyright (2000), with
  permission from Elsevier.}
\label{fig:hybrid_klaus}
\end{center}
\end{figure}

Quark matter in neutron stars has been widely described by using the
MIT bag model with basically one free parameter, the MIT bag constant
$B$.  The onset of the mixed phase from the hadronic to the quark
phase occurs between $(1-2)n_0$ even for large values of the bag
constant $B$ and then sufficiently high densities are reached in the
core of a $1.3M_\odot$ compact star to have quark matter (see
Figure~\ref{fig:hybrid_klaus}). Corrections from hard thermal loop
calculations do not change these numbers significantly.

So called hybrid stars consist of hadronic matter and quark matter and
there are three phases possible: a hadronic phase, a mixed phase and a
pure quark phase. The composition depends crucially on
the parameters as the bag constant $B$ and the interaction strength $g$
between quarks as well as the total mass of the compact star.

In addition, there exists a third solution to the TOV equations
besides the one for the white dwarfs and the one for ordinary neutron
stars, which is stabilized by the presence of a pure quark matter
phase
\cite{Gerlach68,Kaempfer81,Haensel82,Glendenning:1998ag,Schertler:2000xq,Fraga:2001id}.
The third family of compact stars is generically more compact than
ordinary neutron stars and is possible for any first order phase
transition.


\section{Strange Quark Matter in Supernovae}


Stars with a mass of more than 8 solar masses end in a core-collapse
supernova (type II, Ib or Ic). New generation of simulation codes now
have multidimensional treatments and improved neutrino
transport. Still, until recently, no explosions could be
achieved (see e.g.\ \cite{Buras:2003sn}) suggesting missing physics
either with respect to neutrino transport or to the nuclear equation
of state. Only after sufficiently long simulation runs with a quasi
unconstrained geometry a standing accretion shock instability could
develop which leads to an explosion after 600ms \cite{Marek:2007gr}.

The conditions of core-collapse supernova matter at bounce are as
follows: energy densities slightly above normal nuclear matter density
$\epsilon \sim (1-1.5)\epsilon_0$, temperatures of $T\sim 10-20$ MeV
and a proton fraction of $Y_p \sim 0.2-0.3$.
The standard lore for the onset of the quark phase in core-collapse
supernovae is that it happens during the evolution of the
proto-neutron star and not at bounce. The timescale for quark matter to
appear would then be typically $(5-20)$ s after bounce
\cite{Pons:2001ar}, which is, however, due to using rather large bag
constants of $B>180$~MeV. Such a large bag constant would hardly allow
for a pure quark matter phase to develop in the core of cold neutron
stars with a mass of $1.3 M_\odot$ (see
figure~\ref{fig:hybrid_klaus}). The appearance of quark matter would
then be well after the supernova explosion itself. So the question is,
can it be possible to produce quark matter much earlier with a smaller
bag constant?

\begin{figure}
\centerline{\includegraphics[angle=-90,width=0.5\textwidth]{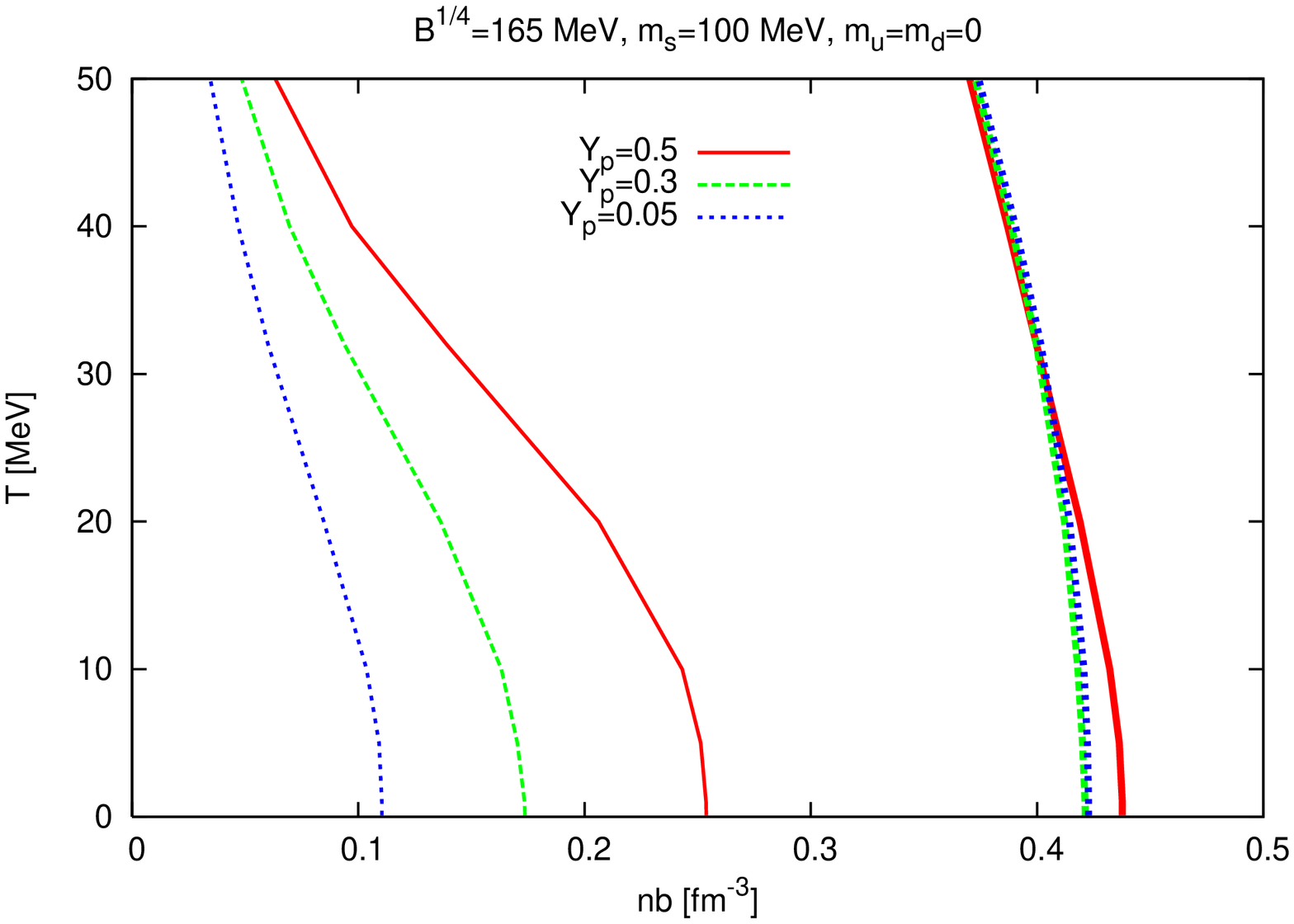}
\includegraphics[angle=-90,width=0.5\textwidth]{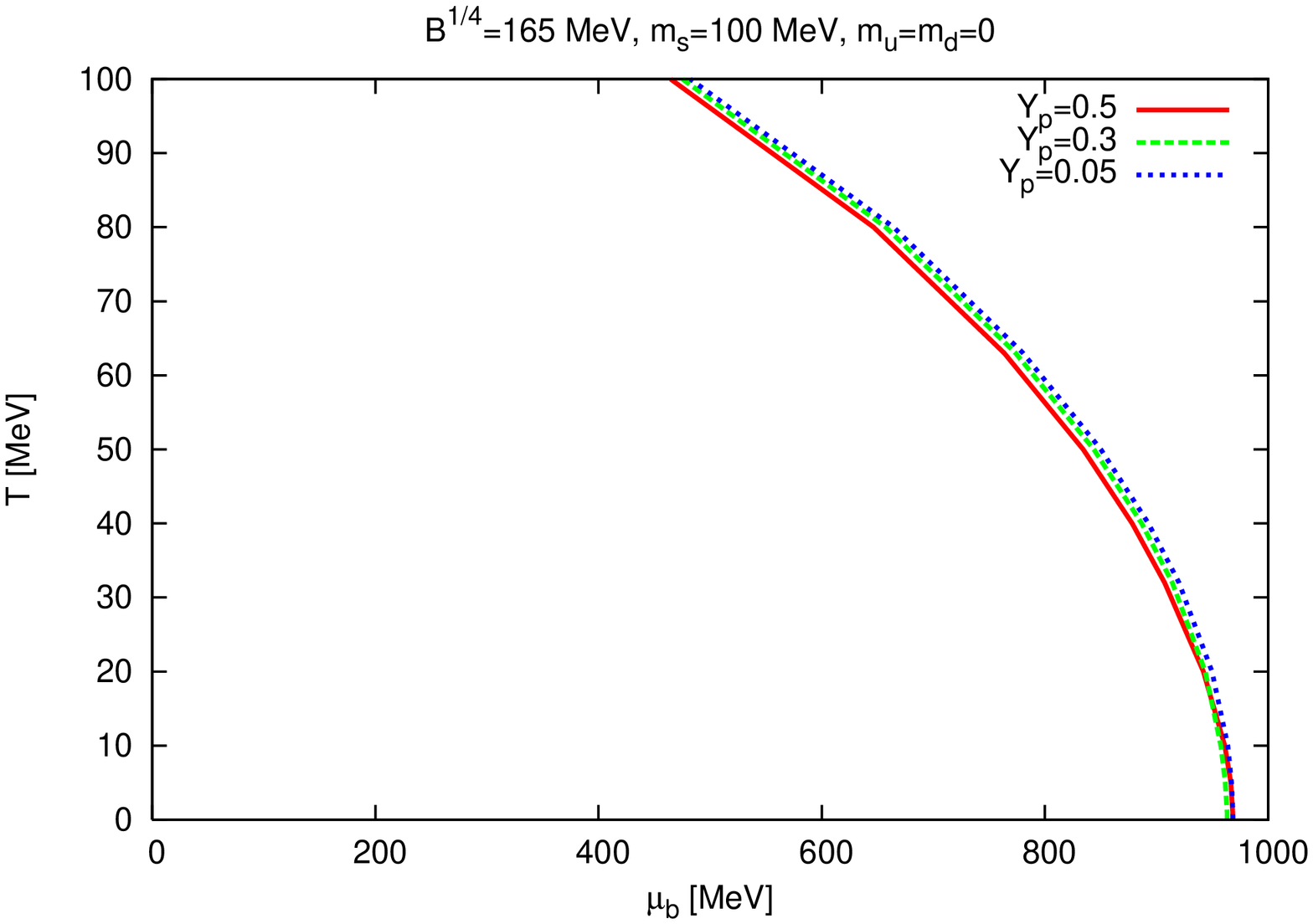}}
\caption{Left plot: The critical density for the phase transition for
  different proton-to-baryon ratios $Y_p$ for supernova conditions;
  thin lines: onset of mixed phase, thick lines: end of mixed phase.
  Right plot: The phase transition line as a function of the baryon
  chemical potential and temperature for supernova conditions.}
\label{fig:crit_astro}
\end{figure}

Figure~\ref{fig:crit_astro} shows the phase transition line to quark
matter for a given temperature versus the net baryon density (left
plot) and versus the baryochemical potential (right plot) for
different values of the proton fraction $Y_p$. The hadronic phase is
modelled by using the EoS of Shen et al.~\cite{Shen:1998gq}, the quark
phase by the MIT bag model with a bag constant of $B^{1/4}=165$~MeV
and a strange quark mass of 100~MeV, the phase transition by a Maxwell
construction.  Interestingly, the critical baryochemical potential is
nearly independent on the proton fraction $Y_p$ and bends towards low
chemical potentials for high temperatures as envisioned in the sketch
of the QCD phase diagram depicted in figure~\ref{fig:qcdphase}.
However, the phase transition occurs at lower densities for lower
proton fractions and higher temperatures, so that neutron-rich hot
supernova material is favourable for the formation of strange quark
matter. Note, that strangeness is not conserved in supernova matter
contrary to the situation in heavy-ion collisions. Strange quark
matter can be formed via the coalescence of hyperons in supernova
matter which are thermally excited in weak equilibrium on timescales
of $10^{-8}$ s or less. Hyperon fractions of about 0.1\% are already
present at bounce, see \cite{Ishizuka:2008gr}.  The lower critical
density in neutron-rich matter is due to the sizable nuclear symmetry
energy so that strange quark matter becomes the energetically
preferred phase.  For supernova material at bounce ($T=10-20$~MeV,
$Y_p=0.2-0.3$, $n\approx n_0$), one reads from the figure that the
immediate production of quark matter is possible!

\begin{figure}
\begin{center}
\includegraphics[angle=-90,width=0.6\textwidth]{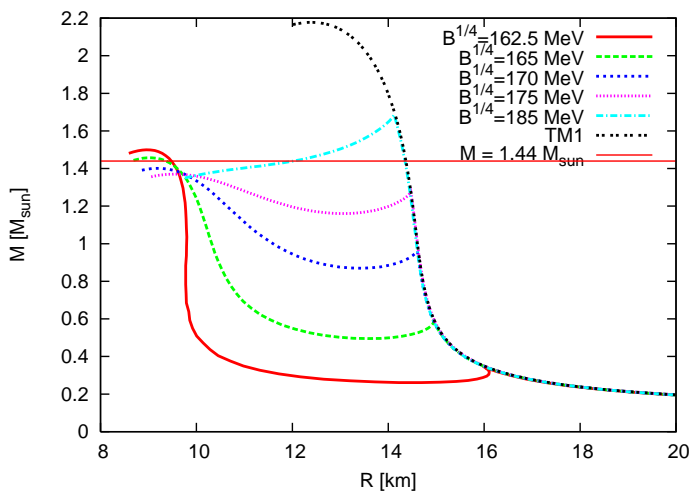}
\caption{Mass-radius relation of cold neutron stars for the supernova
  EoS used for different bag constants. For low bag constants, the
  maximum mass is again above the mass limit from the Hulse-Taylor
  pulsar of 1.44 $M_\odot$ which is shown by the horizontal line.}
\label{fig:mr_sneos}
\end{center}
\end{figure}

Two checks have to be performed in order to be consistent with neutron
star data and heavy-ion phenomenology. First, the maximum mass of a
compact star with the adopted EoS should be at least above $1.44
M_\odot$. The mass-radius relation is plotted in
figure~\ref{fig:mr_sneos} for different values of the MIT bag constant
(for simplicity a Maxwell construction is used for the phase
transition). For large bag constants, the maximum mass is well above
the mass limit from the Hulse-Taylor pulsar. For some intermediate
values, here for the cases of $B^{1/4}=170$~MeV and 175~MeV, the
maximum mass is below $1.44M_\odot$ because of the large jump in
energy density from one phase to the other \cite{Kaempfer81a}. That
was the reason why smaller values of the bag constants were rejected
in the work on proto-neutron stars in
ref.~\cite{Pons:2001ar}. However, for even smaller bag constants of
$B^{1/4}=165$~MeV and below, the maximum mass is again above
$1.44M_\odot$ owing to the increased stability of the pure quark
matter core.  The maximum masses are $1.56 M_\odot$
($B^{1/4}=162$~MeV) and $1.5 M_\odot$ ($B^{1/4}=165$~MeV).  Note, that
the maximum mass for pure quark matter stars, selfbound strange stars,
is well known to be about $2.1M_\odot$ for the original value of the
MIT bag constant, $B^{1/4}=145$~MeV
\cite{Witten:1984rs,Haensel:1986qb,Alcock:1986hz}.

\begin{figure}
\begin{center}
\centerline{
\includegraphics[angle=-90,width=0.5\textwidth]{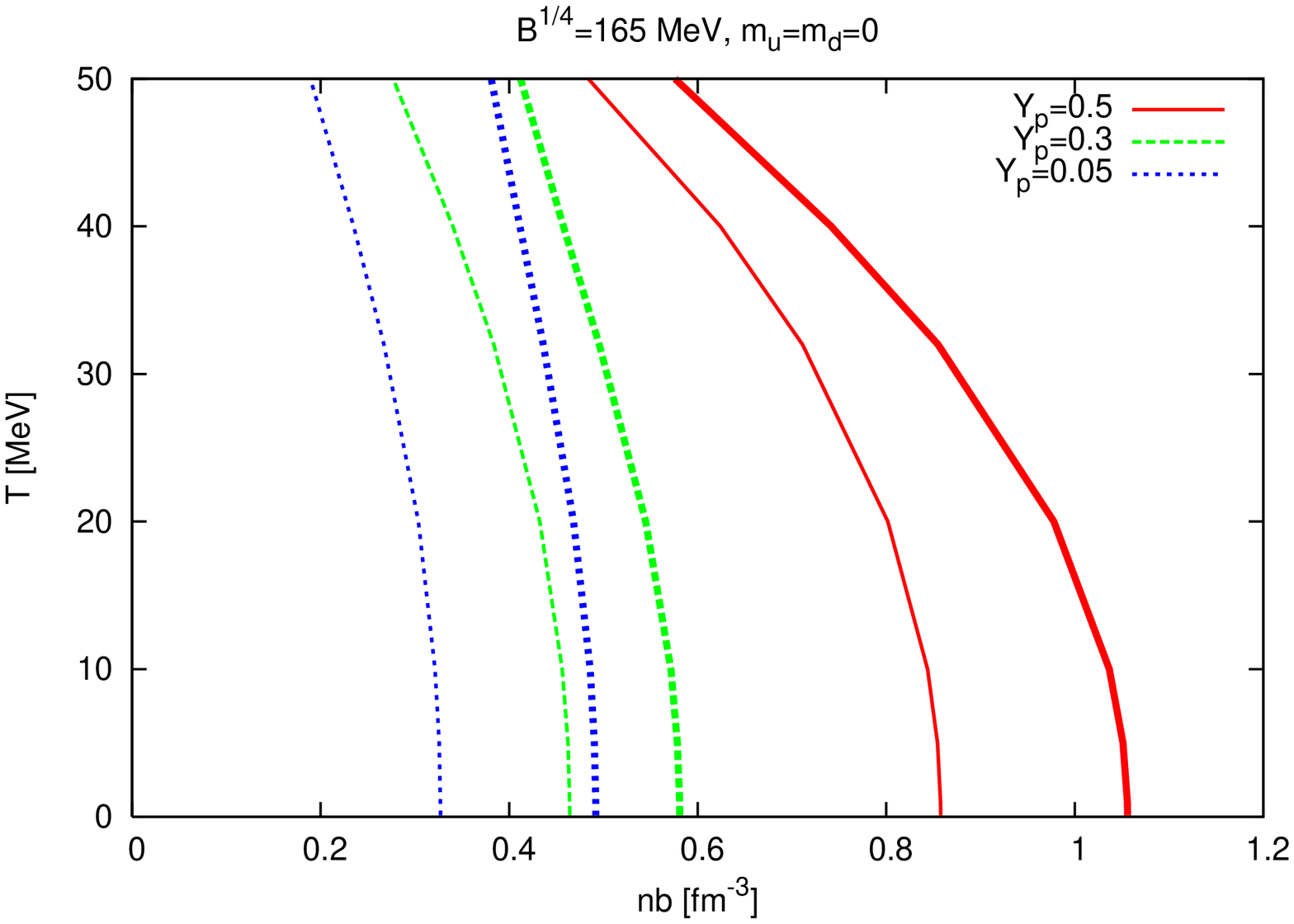}
\includegraphics[angle=-90,width=0.5\textwidth]{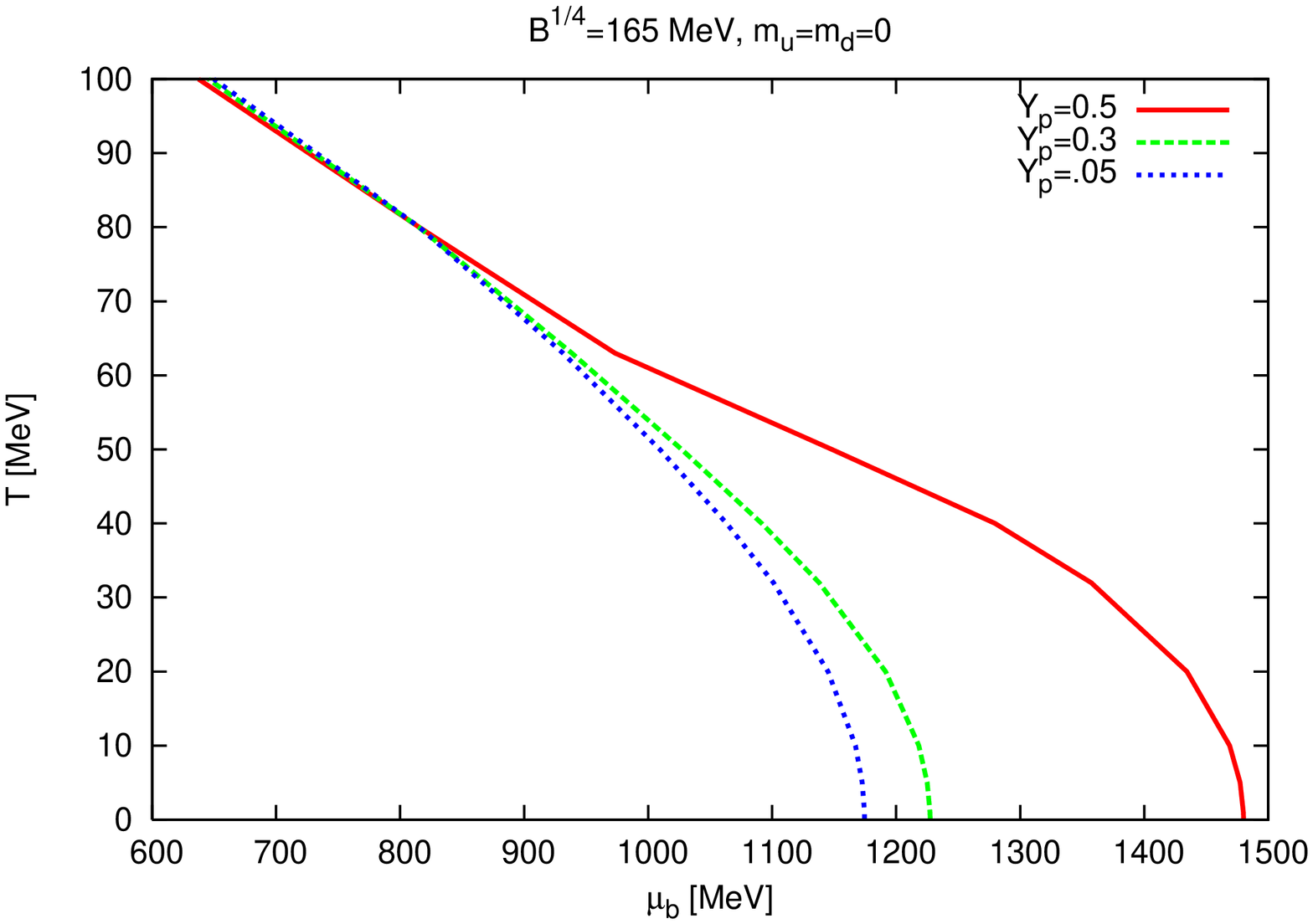}}
\caption{Left plot: The critical density for the phase transition
  lines for different proton-to-baryon ratios $Y_p$ for heavy-ion
  conditions (no weak equilibrium for strangeness); thin lines:
  onset of mixed phase, thick lines: end of mixed phase.\\
  Right plot: The phase transition line as a function of the baryon
  chemical potential and temperature for heavy-ion conditions.}
\label{fig:crit_hic}
\end{center}
\end{figure}

Secondly, we have to calculate the phase transition line appropriate
for matter produced in relativistic heavy-ion collisions. The
timescales are too short to generate weak equilibrium in heavy-ion
collisions, so the initially produced quark matter has net strangeness
zero (of course thermal production of strange quark pairs is possible
but highly suppressed for the low temperatures relevant for supernova
explosions).  Therefore, the quark matter formed consists mainly out
of pure up- and down-quarks and becomes highly unfavoured compared to
the case of supernova matter, where strange quark matter is formed.
Indeed, we find large critical densities for the phase transition, in
particular for isospin-symmetric matter (proton fraction $Y_p=0.5$)
which is relevant for the heavy-ion case. The phase transition occurs
at much larger baryon densities, well above five times normal nuclear
matter density for low temperatures so that low-energy heavy-ion
collisions can not produce quark matter.  Also the phase transition
line is located at larger chemical potentials for up-down-quark
matter. The location of the freeze-out points in statistical
approaches is about $\mu_{f.o.}= 700-800$~MeV, $T_{f.o.}=50-70$~MeV
for SIS energies and $\mu_{f.o.} \sim 500$~MeV, $T_{f.o.}\sim 120$~MeV
for AGS energies.  These values are lying below the phase transition
line in the corresponding right plot of fig.~\ref{fig:crit_hic}. Note,
that the hadronic equation of state has been modelled to suit its
applications for supernova simulations. Therefore, temperatures of
more than 50~MeV can not be handled appropriately in principle, but
are also not relevant for our purposes here. For the application to
higher temperatures, one needs to improve the hadronic equation of
state by e.g.\ incorporating pions, kaons, hyperons and resonances
which would shift the phase transition line to even higher densities
and baryochemical potentials.

Now as the supernova EoS has passed these two tests, let us discuss the implications for core-collapse supernovae \cite{Sagert:2008ka}.
The stiffening of the nuclear EoS above normal nuclear matter densities
produces a bounce of the supernova material, a shock wave is moving outwards but stalls around 100 km. Afterwards, quark matter is formed in the core and at a few 100ms the mixed phase collapses and an accretion shock develops on the pure quark matter core. The accretion shock turns into an outgoing shock wave which is so energetic that it runs over the stalled first shock leading
to an explosion! Note, that normally 1D supernova simulations are not able to achieve an explosion, only multidimensional codes are presently capable of producing an explosion with the help of the standing accretion shock instability, see \cite{Marek:2007gr}. 

Our supernova simulation runs were performed for different parameters,
where the quark core appears at $t_{\rm pb}=200$~ms to $500$~ms post bounce.
The results ($t_{\rm pb}$, baryonic mass and explosion energy) are
significantly sensitive to the location of the QCD phase transition
(i.e.\ the bag constant in our case). Heavier progenitor masses can lead to the formation of a black hole which could be circumvented by stiffening the quark EoS in order to explain the rather long emission of neutrinos from SN1987A.

\begin{figure}
\begin{center}
\includegraphics[width=0.5\textwidth]{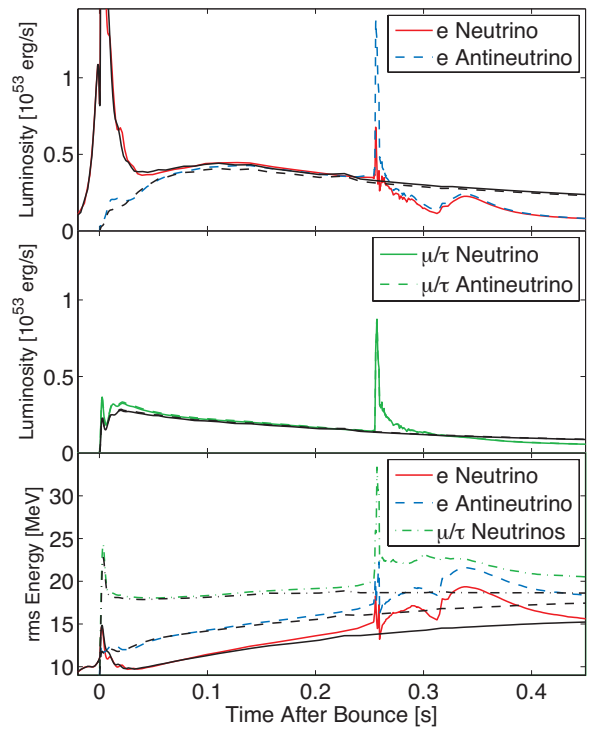}
\caption{The neutrino spectrum without a phase transition (thick
  lines) and with a phase transition (thin lines). The case with a
  phase transition to strange quark matter results in a second peak in
  antineutrinos. The average energies of the emitted neutrinos
  increases also. Reprinted figure with permission from
  \cite{Sagert:2008ka}. Copyright (2009) by the American Physical
  Society.}
\label{fig:sn_neutrinos}
\end{center}
\end{figure}

Most interestingly, we find that the temporal profile of the emitted neutrinos out of the supernova reflects the features of the QCD phase transition.
Figure~\ref{fig:sn_neutrinos} shows the neutrino luminosity and the mean energy as a function of time. The first peak in electron neutrinos is due to the first shock. When the QCD phase transition is included we find a second peak in electron {\em anti}-neutrinos at about the time when the strange quark matter core is created. 
The pronounced second peak of anti-neutrinos is due to the protonization of the material when the second shock front runs over the neutrinosphere.
We note that the location of the second peak and its height is controlled by the critical density and strength of the QCD phase transition!


\section{Summary}


The QCD phase transition to strange quark matter leads to a rich
variety of astrophysical signals involving compact stars and
supernovae.  Neutron stars with a core of strange quark matter are
compatible with present neutron star data. Strange quark matter can be
formed in supernovae, even shortly after the first bounce. A second
outgoing shock is generated which has enough energy to lead to an
explosion.  The presence of a strong QCD phase transition
can be read off from a second peak in the (anti-)neutrino signal.  The
formation of strange quark matter will certainly have also significant
implications for the gravitational wave signal of core-collapse
supernovae and for the r-process nucleosynthesis as core-collapse
supernovae are considered to be the prime astrophysical site. 

\ack 

This work is supported by the German Research
Foundation (DFG) within the framework of the excellence initiative
through the Heidelberg Graduate School of Fundamental Physics,
the Gesellschaft f\"ur Schwerionenforschung
mbH Darmstadt, Germany, the Helmholtz Research School for Quark
Matter Studies, 
the Helmholtz Alliance Program of the Helmholtz Association, contract HA-216 "Extremes of Density and Temperature: Cosmic Matter in the Laboratory",
the Frankfurt Institute for Advanced Studies,
the Italian National Institute for Nuclear Physics, the Swiss National
Science Foundation under the grant numbers PP002-106627/1 and
PP200020-105328/1, and the ESF CompStar program.
A.M. is supported at the Oak Ridge National Laboratory, managed by
UT-Battelle, LLC, for the U.S. Department of Energy under contract
DE-AC05-00OR22725.


\section*{References}

\bibliographystyle{utphys}
\bibliography{all,literat}

\end{document}